\documentclass{article}

\usepackage[final]{NeurIPS_2019}

\usepackage[utf8]{inputenc} 
\usepackage[T1]{fontenc}    
\usepackage{hyperref}       
\usepackage{url}            
\usepackage{booktabs}       
\usepackage{amsfonts}       
\usepackage{nicefrac}       
\usepackage{microtype}      
\usepackage{here}
\usepackage{graphicx}
\usepackage{color}

\title{Towards Federated Graph Learning for Collaborative Financial Crimes Detection}

\usepackage{authblk}
\author[1]{Toyotaro Suzumura}
\author[2]{Yi Zhou}
\author[2]{Nathalie Baracaldo}
\author[1]{Guangann Ye}
\author[1]{Keith Houck}
\author[3]{Ryo Kawahara}
\author[2]{Ali Anwar}
\author[4]{Lucia Larise Stavarache}
\author[3]{Yuji Watanabe}
\author[3]{Pablo Loyola}
\author[1]{Daniel Klyashtorny}
\author[2]{Heiko Ludwig}
\author[1]{Kumar Bhaskaran}
\affil[1]{IBM T.J. Watson Research Center}
\affil[2]{IBM Research Almaden}
\affil[3]{IBM Research Tokyo}
\affil[4]{IBM Global Business Services}
{
    \makeatletter
    \renewcommand\AB@affilsepx{: \protect\Affilfont}
    \makeatother

    \makeatletter
    \renewcommand\AB@affilsepx{, \protect\Affilfont}
    \makeatother

}

\begin{document}

\maketitle

\begin{abstract}

 Financial crime (e.g., fraud, theft, money laundering) is a large and growing problem, in some way touching almost every financial institution, as well as many individuals, and in some cases, entire societies.  Financial institutions are the front line in the war against financial crime and accordingly, must devote substantial human and technology resources to this effort. Current processes to detect financial misconduct (including the technologies used) have limitations in their ability to effectively  differentiate between malicious behavior and ordinary financial activity. These limitations tend to result in gross over-reporting of suspicious activity (typically manifested as "alerts") that necessitate time-intensive and costly manual review. Advances in technology used in this domain, including machine learning based approaches, can improve upon the effectiveness of financial institutions' existing processes, however, a key challenge that most financial institutions continue to face is that they address financial crimes in isolation without any insight from other firms. Where financial institutions address financial crimes through the lens of their own firm, perpetrators may devise sophisticated strategies that may span across institutions and geographies. Financial institutions continue to work relentlessly to advance their capabilities, forming partnerships across institutions (including governmental bodies) to share insights, patterns and capabilities. These public-private partnerships are subject to stringent regulatory and data privacy requirements, thereby making it difficult to rely on traditional technology solutions. In this paper, we propose a methodology to share key information across institutions by using a federated graph learning platform that enables us to build more accurate machine learning models by leveraging federated learning and also graph learning approaches. We demonstrated that our federated model outperforms local model by 20\% with the UK FCA TechSprint data set. This new platform opens up a door to efficiently detecting global money laundering activity.  
 

\end{abstract}

\section{Introduction}

\subsection{Financial Crimes} 

Financial crime ~\citep{aml-chen-2018} \citep{NextGenAML2018} \citep{aml-jamshidi-2019} \citep{aml-alexandre-2018} \citep{Savage2016DetectionOM}
\citep{FRONZETTICOLLADON201749}

is a broad and growing class of criminal activity involving the misuse, misappropriation, or misrepresentation of entities with monetary value.  Common subclasses of financial crime include theft, fraud, and money laundering (i.e., obscuring the true origin of monetary entities to evade regulations or avoid taxes).  The monetary value of such crimes can range from tens of dollars to tens of billions of dollars, however, the overall negative consequences of such crimes extend far beyond their monetary value.  In fact, the consequences may even be societal in scope, such as in cases of terrorist financing or large-scale frauds that topple major institutions and governments (e.g., mortgage crisis, 1MDB scandal).
In response, financial institutions spend substantial resources to develop compliance programs and infrastructures in order to combat financial crime. Managing financial crime risk presents challenges due to the scale of the effort (large banks may have upwards of 100M customers or more, which together generate billions of transactions that must be screened) and availability of data (when transactions cross bank or country boundaries, little may be known about the remote counterparty). Current technology employed to assist with these processes focuses on the identification of anomalies and/or known patterns malfeasance, but usually also generates a large number of false positive alerts in the process. These alerts then require further (often manual) review to parse out suspicious behavior from valid financial activity that is inadvertently picked up by the models (referred to as "false positives").


\subsection{Combating Financial Crimes with Machine Learning and Graph Learning}

Recently, financial institutions have been exploring the use of machine learning techniques to augment existing transaction monitoring capabilities. Machine learning techniques offer a promising capability to identify suspicious activity from an incoming stream of transactions, as well as to filter the false positives from the alerts generated by current technology, thereby making existing processes more efficient and ultimately more effective. These machine learning techniques rely on a set of features generated from knowledge about the transacting parties, from individual and aggregate transaction metrics, and from the topology of party-to-party relationships derived from static knowledge and transactional history.  Topological features are computed from network embeddings or from the results of traditional graph algorithms
(e.g., PageRank, count of suspicious parties within an egonet in the following figures ). Overall, this approach has been shown to have a positive effect when evaluated against a ground truth determined by currently deployed methods. In one such evaluation, false positives were reduced 20-30\%.  

\subsection{Needs for Global Financial Crimes Detection and  Contributions}

Notwithstanding the value of leveraging machine learning in the context of transaction monitoring, financial institutions are limited to identifying  suspicious activity as it pertains to their organization. This presents a conundrum since bad actors are increasingly sophisticated with their techniques that often span across organizations and geographies (i.e., many use multiple banks to launder money). Financial institutions are realizing that without looking at data across multiple organizations, it would be impossible to detect a portion of suspicious activity. Regulatory requirements, data privacy concerns, as well as commercial competitiveness, all pose challenges to explicit sharing in information amongst financial institutions. Given the challenge at hand, a innovative solution is needed to detect suspicious activity across organizations. Our contributions are the following: \\ 

\begin{itemize}
\item Developed a federated graph learning platform that detects global financial crime activities across multiple financial institutions\
\item Demonstrated the effectiveness of federated graph learning as a tool to help identify financial crime during the TechSprint hosted by the Financial Conduct Authority earlier this year, using the data set and use cases provided by the FCA. 
\item First to demonstrate the ability to combine federated learning with graph learning as a means to detect potential financial crimes and share typologies across multiple financial institutions. 
\end{itemize} 

The rest of this paper is organized as follows. We outline our core technologies, together with related work, followed by the overall architecture and our federated graph learning capabilities. We then provide an overview of preliminary implementation and evaluation using the data set provided by UK FCA TechSprint. Finally we describe concluding remarks and future directions.

\section{Related Work}

This section describes the underlying technologies, along with relevant prior art, that constitute our platform - including graph-based machine learning and federated learning.

\subsection{Graph-based Machine Learning}
Graph learning is defined as a type of machine learning that utilizes graph-based features to add richer context to data by first linking that data together as a graph structure, and then deriving features from different metrics on the graph. Various graph features can be defined by exploiting a set of graph analytics such as connectivity, centrality, community detection, pattern matching. Graph features can also be combined with non-graph features (e.g., features on attributes for a specific data point). Once a set of features including graph features and non-graph features are defined, a problem can then be formulated as a supervised machine learning problem (assuming that label data is provided). However, if label data is not provided, it would be an unsupervised machine learning problem so that we can apply clustering (e.g. kmeans) or outlier detection (e.g. LoF or DBScan). Recently, there have been many advances in scalable graph computation for billion-scale or even trillion-scale graphs\citep{graph500,DBLP:journals/corr/abs-1908-05855}, so
it is reasonable to expect that this approach would remain practical, even for large graphs.

The paper in ~\citep{Akoglu-2015-graph} provides a good review of prior art about graph-based approaches for general anomaly detection problems. \citep{ian-2016-fraud} uses PageRank-based features for fraud detection. We are also exploring graph embedding methods for financial crime detection applications such as anti-money laundering (AML) in \citep{DBLP:journals/ijon/LiuLYCSH19} and \citep{DBLP:journals/corr/abs-1812-00076}, but we are still in the process of applying those methods to real world financial data sets. Recently communities are also exploring the use of neural networks to compute graph embeddings without determining pre-defined graph topologies as graph features. However, the lack of explainability of the black-box model of neural network presents adoption challenges for financial institutions that have stringent model validation processes that hinge on explainability of the decisions made and outputs that result. 
 
Notably these prior works focus more on local graph features, while our current work is focused on global graph features spanning multiple financial institutions.

\subsection{Federated Learning }

Traditional machine learning requires all the training data to be collected and accessible to a trusted third party. However, privacy concerns and legislation such as General Data Protection Regulation (GDPR) and Health Insurance Portability and Accountability Act (HIPAA) has presented many challenges to being able to transmit data to a centralized location for training. 

As a result, "federated learning" \citep{fl-mcmahan-2017,fl-sysml09,fl-2016,fl-bonawitz-2016} has emerged as an alternative way to do collaborative model training without sharing the training data. In Federated Learning, each data owner maintains its own data locally and engage in a collaborative learning where only model updates, such as model parameters, are shared.

Examples of federated learning scenarios include a large number of individual parties providing personal data to smart phone apps and a relatively small number of competing companies within the same domain training a single model \citep{fl-mcmahan-2017}. As of this writing, no one has pursued leveraging federated learning in the context of mitigation of financial crimes risk.








\section{Federated Graph Learning for AML}

We propose a new platform that enables us to capture complex global money laundering activities spanning multiple financial institutions as opposed to current AML systems that only look at transactions at single bank. The proposed federated graph learning system is comprised of 3 steps, we compute local features, compute global graph features, and then perform federated learning over computed features. Subsequent sections describe each step. 


\subsection{Local Feature Computation}
We firstly compute local features for each financial institution. As local features, we can firstly compute demographic features of customers such as account types (individual or business), business types, countries, account opening date, and some risk flags based on "know your customer" (KYC) attributes. We then compute various statistical features on transaction behaviors such as min, max, average, mean, standard deviation for transaction of various types such as international wire, domestic wire, credit, cash, check, and so forth. We can also compute graph features such as egonet, pagerank, degree distribution as what have done for a single bank case.   


\subsection{Global Feature Computation}
As a next step, we compute global features that provide global context related to suspicious activities among multiple financial institutions. Global graph features are mainly computed using graph analytics over the entire graph of transaction and party relationship graph.
These include 1 hop / 2 hop egonets, cycle and temporal cycle, betweeness centrality, community detection, and so forth. The advantage of using global features over local graph features is if we can create richer and denser graph by assembling sub-graphs from multiple graphs, then the graph features should be more effective since you can also acquire contexts from other financial institutions as to which bank accounts may be associated with bad actors.


In computing global graph features, we need to take privacy into account, so as not to disclose any sensitive information from each financial institution. For instance, if there is a cycle of transactions consisting of 3 accounts in two different financial institutions - starting from an account A in Financial Institution X to an account B in Financial Institution Y, and to an account C in Financial Institution Y. A challenge is that a transaction from B to C in Financial Institution Y cannot be revealed to Financial Institution X. Thus, one of the requirements is to design and implement a secure protocol that allows Financial Institution X to send a inquiry to Financial Institution Y to ask whether there is a transaction between B and C - without letting Financial Institution Y to reveal sensitive information. GraphSC \citep{graphsc2015} is one of such secure graph computation frameworks, and we have started to implement some graph features such as temporal cycle features.










\subsection{Federated Learning}

Next, we build a federated machine learning model using local features and global features that we describe in the previous sections using our federated learning framework \citep{fl-almaden-2018}.  For this work, we use our centralized federated learning in which data owners share model updates with a central server, aka an aggregator\footnote{Note that the aggregator does not have access to the data of any of the parties.}. This central server is hosted by a third-party such as Financial Intelligence Unit~(FIU). To further protect the privacy, even the model updates shared with the aggregator are strictly secured via privacy preserving techniques such as differential privacy, multi-party computation, and / or an array of encryption techniques.

We target a scenario where different Financial Institutions collaborate together to train a model that can more accurately predict suspicious money laundering efforts. In our setup, each Financial Institution trains on its local data and shares the model parameters of the trained model with the central aggregator. The aggregator then fuses all of the model parameters and generates a global model whose weights will be sent back to all the collaborating banks to reinitialize their local model for another round of local training. This process is repeated for a set number of rounds or until desired model accuracy is achieved.


Our framework for federated learning (FFL)\citep{fl-almaden-2018} is a framework designed for federated learning in an enterprise environment. It provides a basic fabric for federated learning on which advanced features, such as, differential privacy and secure multiparty computation, can be added. It is agnostic to the specific machine learning platform used and supports different learning topologies, e.g., a shared aggregator, and protocols. 
Different from Google TensorFlow Federated (TFF) and Openminded PySyft, FFL not only supports simulated federated learning environments but also real distributed learning environments with different connection configurations. Moreover, FFL enables training a variety of machine learning models, e.g., decision tree, neural networks on keras, linear models on scikit-learn, in a federated learning fashion.

\section{Preliminary Implementation and Evaluation }

In this section we describe a work-in-progress prototype implementation for our federated graph learning framework described in the previous section. For the evaluation described in this paper, we did not use a secure graph computation framework such as \citep{graphsc2015} to compute global graph features. Secure graph computation is still work in progress, to be described in a future publication.

\subsection{Data Set and Graph Modelling} 

For the evaluation described in this paper, we used the data set provided by the FCA during the 2019 TechSprint \citep{ukfca-techsprint2019}. The data set is a simulated data set comprised of data from 6 financial institutions in the UK and reflects real-world statistical distributions and well-known suspicious patterns. 

The data set spanned 2 years of activity and includes customer profile, transactions, customer relationship data, an indicator of suspicious activity alerts, and an indication of whether the customer relationship was terminated over suspicion of misconduct. We used the last item as a form of ground truth for suspicious activity.  

On the basis of this data set, we built two types of graphs, one called a transaction graph where a vertex represents a bank account and an edge represents a money transfer. Another graph is called party relationship graph where a vertex represents a customer and an edge represents a social relationship between customers such as family. 


  


%

\subsection{Graph Features for Party Relationship Graph}

Due to space limitations, we focus here exclusively on the party relationship graph which consists of social relationships between bank accounts. For example, a person who owns a company can use both his or her personal account and the company's business account. In this case, those two bank accounts can be related through the owner. This kind of relations could be an important indicator of a financial crime because a criminal might use an account indirectly through the relationship (e.g., ownership of a company) to send his or her private money to obscure the true source or beneficiary (i.e., the layering).

Here, we assume that a bank has the following information for each customer:

\begin{itemize}
    \item customer profile (e.g., account ID, name, date of birth, nationality, etc.),
    \item related party profiles (e.g., name, date of birth, etc.), 
    \item relations between the customer and the related parties (e.g., director, owner, family, etc.), and
    \item customer risk intelligence (e.g., past Suspicious Activity Report (SAR) flags, financial crime exit markers).
\end{itemize}
Such information is obtained during the KYC processes (when onboarding new customers or in performing periodic reviews of existing customers) or when performing detailed investigations of AML alerts. The related parties may or may not be a customer of a financial institution, and could include the customer itself.
If multiple customers have relationships with a common related party, this indicates that the accounts of the customers might be affected by a single party and thus could work in a coordinated manner.

Since there are many financial institutions in the market, one needs to consider the case of an individual having accounts in multiple financial institutions. Similarly, the same related party could appear in the data of multiple financial institution. To reveal the connection between accounts across the financial institution boundaries, one needs to go through the process of "entity resolution" to draw the connections between the customer profiles and related party profiles. That is, one needs to identify the profiles which correspond to a same entity by comparing the attributes such as the names, addresses, or identification numbers.

Here, we applied the following simple rule for the entity resolution.
\begin{itemize}
    \item Individual customer or party:
    (full name, date of birth, and nationality are equal) OR
    (ID document type, ID document number, and nationality are equal)
    \item Business customer or party: 
    (full name, date of incorporation, and country of incorporation are equal) OR
    (company registration type, company registration number, and country of incorporation are equal)
\end{itemize}
However, in practice, entity resolution presents many challenges due to the existence of typos, document quality issues, OCR errors, or fluctuations in conversions of non-Latin characters. There are a number of commercial products that address this challenge in contexts where raw data can be shared, however, performing entity resolution under privacy preserving constraints remains an area of future work.

\begin{figure}[tb]
 \begin{center}
  \includegraphics[clip, width=0.5\hsize]{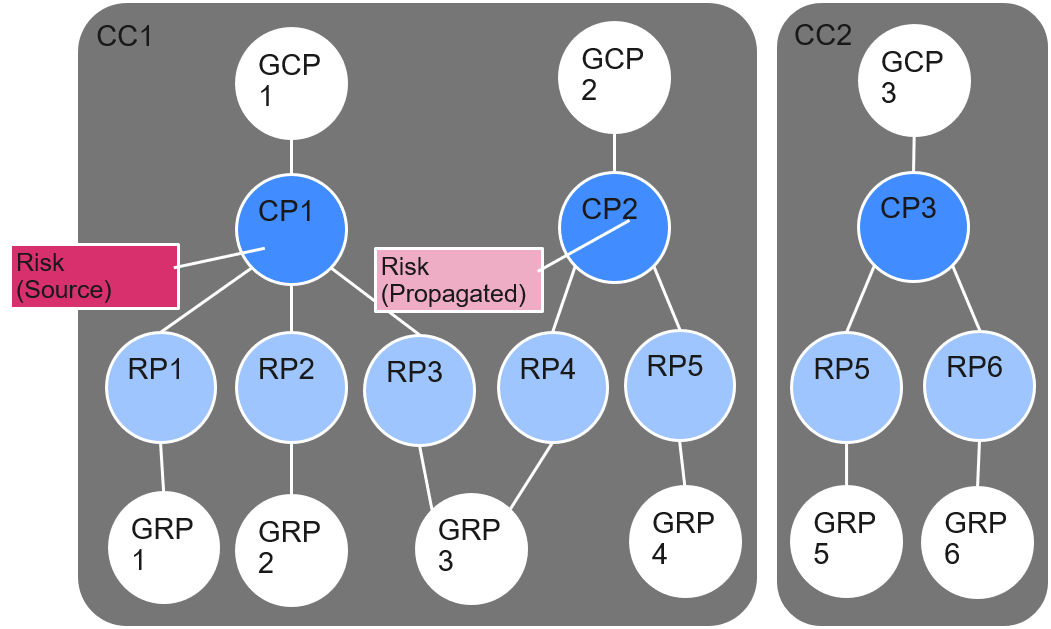}
  \caption{Relation between customers and related parties.}
  \label{fig:ryokawa-customer-relation}
 \end{center}
\end{figure}

Once the entity resolution for the customers and their related parties has been performed, one will get a graph of those entities, as shown in Fig. \ref{fig:ryokawa-customer-relation}.
In the figure, CP1, CP2, $\cdots$ are the customers, each of which has an account,
RP1, RP2, $\cdots$ are their related parties, GRP1, GRP2, $\cdots$ and GCP1, GCP2, $\cdots$
are the grouping IDs issued during the entity resolution.
Edges between the customers and the related parties are the social relations, and
edges between the grouping IDs and the customers or related parties are created
if those parties are identified as belonging to the same entity by the entity resolution.

Since the connected accounts (customers) in the graph are possible collaborators,
we think that the risk of being involved in money laundering is shared among the accounts.
From this hypothesis, we compute each customer's features based on the statistics in
each (weakly) connected component in the graph. 
In the current implementation, the following features are used:
\begin{itemize}
    \item number of customers who have alerted by a transaction monitoring system in the past within the connected component.
    \item number of customers who have SAR flags in the past within the connected component,
    \item number of customers who have financial crime exit markers in the past within the connected component,
    \item number of the nodes within the connected component.
\end{itemize}
The status of the risk flags (SAR, financial crime exit marker, etc.) can be obtained from
the customer risk intelligence data as mentioned in a previous paragraph in this section.
Please note that the risk flag status of a customer is often used as
a target variable in a machine learning-based prediction / classification task of financial crimes. In such cases, those features must not include
the status of the risk flag of the customer in question and 
must contain the information from only other customers
when those are used as a training or a testing data set.

\begin{figure}[tb]
 \begin{center}
  \includegraphics[clip, width=0.45\hsize]{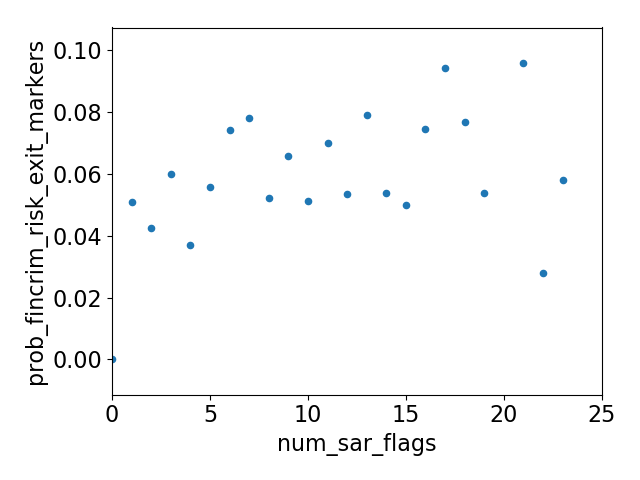}
  \includegraphics[clip, width=0.45\hsize]{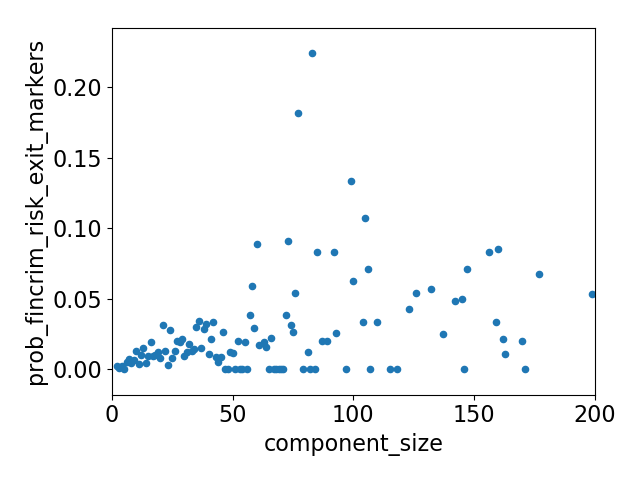}
  \caption{Conditional probability of a customer's fincrime exit marker being flagged as a function of a feature. 
  Left: the feature is the number of SAR-flagged customers in 
  the same connected component. 
  Right: the feature is the number of nodes in the connected component.}
  \label{fig:ryokawa-summary-sar}
 \end{center}
\end{figure}

Our preliminary analysis on the synthetic data set used during the TechSprint
is shown in Fig. \ref{fig:ryokawa-summary-sar}.
Here, we assume that the financial crime exit marker of a customer is the target variable to be predicted for supervised machine learning setting. 
It shows a positive correlation between the probability of a customer
being flagged with fincrime exit marker and 
the number of customers who have SAR flags (left) within the same
connected component and
the number of nodes (including the customers, related parties and the grouping IDs)
within the same connected component.
This result indicates that these values can be used as features for
detecting money laundering with other features.


\subsection{Evaluation of Federated Graph Learning}

Here we show the evaluation result of our federated graph learning using the TechSprint data. We compute local features including transaction-based features, and compute global graph features in party relationship graph defined in the previous section, and global graph features in transaction graph, and local transaction features.  

With regards to the platform setting, in an ideal federated learning environment, each financial institution will perform its local training on its own server or virtual machine and communicate with the aggregator, which can be hosted by a third party (e.g., government agency) or by one of the banks, after each local training period is performed. 
However, due to the limited resources provided by during the TechSprint, we only had one host, and hence needed to simulated 6 processes representing 6 UK banks' local training processes and one process was used as a proxy for the role of the aggregator using our federated learning framework \citep{fl-almaden-2018} 
We have tried to train several types of machine learning models, for example, $\ell$-1 regularized logistic regression,  $\ell$-2 regularized linear support vector machine (SVM), decision tree and a simple neural network. 

We found similar performance results (with less than 10\% difference in testing accuracy and F1 scores) for these machine learning models. Therefore, we only report the results for neural network, which composed of two dense layers of sigmoid units and a sigmoid layer with binary cross-entropy loss. As previously noted, explainability for neural networks is still an ongoing research area, so we could leverage existing work or use other machine learning models for the current financial regulation policy that requires transparency and explainability in the machine learning models that are used. 
Since the dataset that was provided obtained is highly imbalanced with only around 5\% bank accounts containing labels on whether they were filed as Suspicious Activity Reports (SAR) and around 0.4\% are labeled as financial criminals, we exploited the under-sampling strategy in the majority label class, (i.e., the clean bank accounts, to create balance training datasets for all financial institutions). 
We then trained local models and the aggregated model all based on the balanced training datasets. 

In Table~\ref{tab:local-trans}, we provide the results of local models trained on each financial institution's transaction records.
We observed that since the local test sets are balanced, the test accuracy and F1 score are the same, which seems to demonstrate good performance of the local models. 
However, if we test the local trained model against account records from all financial institutions, we can see that F1 scores drop significantly due to the imbalanced nature of the test set and the test accuracy also drops a bit.
Moreover, if we add graph features that we described in the previous section into our training features, we see improvements in both test accuracy and F1 scores as shown in Table~\ref{tab:local-trans-graph}.
From Table~\ref{tab:FL-model}, we can conclude from the results that by training an aggregated model collaboratively via federated learning, all financial institutions can benefit from the aggregated model without sacrificing their data privacy.  
\begin{table}[]
\centering
\resizebox{\textwidth}{!}{%
\begin{tabular}{|l|l|l|l|l|l|l|}
\hline
&BWBAGB & PCOBGB & NUBAGB & HCBGGB & GVBCGB & FOCSGB \\ \hline
\begin{tabular}[c]{@{}l@{}}Local test set \\ (Accuracy/F1)\end{tabular} & 0.971/0.971 & 0.976/0.976 & 0.984/0.984 & 0.982/0.982 & 0.967/0.966 & 0.988/0.988 \\ \hline
\begin{tabular}[c]{@{}l@{}}All record test set\\ (Accuracy/F1)\end{tabular} & 0.956/\textcolor{red}{0.550} & 0.956/\textcolor{red}{0.550} & 0.953/\textcolor{red}{0.546} & 0.957/\textcolor{red}{0.551} & 0.960/\textcolor{red}{0.550} & 0.962/\textcolor{red}{0.552} \\ \hline
\end{tabular}%
}
\caption{Centralized local models trained on transaction features}
\label{tab:local-trans}
\end{table}

\begin{table}[]
\centering
\resizebox{\textwidth}{!}{%
\begin{tabular}{|l|l|l|l|l|l|l|}
\hline
 & BWBAGB & PCOBGB & NUBAGB & HCBGGB & GVBCGB & FOCSGB \\ \hline
\begin{tabular}[c]{@{}l@{}}Local test set \\ (Acc/F1)\end{tabular} & 0.996/0.996 & 0.997/0.997 & 0.997/0.997 & 0.996/0.996 & 0.990/0.990 & 1/1 \\ \hline
\begin{tabular}[c]{@{}l@{}}All record test set\\ (Acc/F1)\end{tabular} & 0.994/\textcolor{red}{0.761} & 0.994/\textcolor{red}{0.769} & 0.990/\textcolor{red}{0.692} & 0.995/\textcolor{red}{0.766} & 0.995/\textcolor{red}{0.764} & 0.995/\textcolor{red}{0.765} \\ \hline
\end{tabular}%
}
\caption{Local models trained on transaction and graph features}
\label{tab:local-trans-graph}
\end{table}

\begin{table}[H]
\centering
\begin{tabular}{|c|c|}
\hline
 & Aggregated model \\ \hline
Accuracy & \bf 0.995 \\ \hline
F1 & \bf 0.769 \\ \hline
\end{tabular}
\caption{Federated model trained on transaction and graph features}
\label{tab:FL-model}
\end{table}




\section{Concluding Remarks and Future Directions}
In this paper we proposed a novel framework that enables us to better identify patterns of suspicious activity by sharing insights across  multiple financial institutions without sharing any raw data from each financial institution. This was made possible by combining graph-based machine learning techniques with federated learning (referred to as, federated graph learning). We described the overall architecture, work-in-progress implementation, demonstrated that the federated graph learning model using multiple financial institutions outperforms local model by 20\% based on a data set from the 2019 FCA TechSprint. We believe that this capability lays the foundation for us to be able to pilot these techniques on real world data and scenarios. In order to do so, we are actively working on a pilot where multiple financial institutions participate in utilizing federated graph learning to supplement their existing financial crime mitigation framework. In designing the pilot, we are exploring the roles that can be played by financial institutions, regulators, FIUs, technology firms and consultancies to achieve maximum results.  


%
\subsubsection*{Acknowledgments}
The use cases and computing environment were provided by UK FCA (Financial Conduct Authority) TechSprint in August 2019. The data was also provided by Harbr during TechSprint. 


\newpage
\bibliographystyle{plainnat}
\bibliography{reference}

\end{document}